\documentstyle[ltwol,psfig]{article}

\def\be{\begin{equation}}
\def\ee{\end{equation}}
\def\bea{\begin{eqnarray}}
\def\eea{\end{eqnarray}}

\begin{document}

\title{Charmless $B$ Decays Involving  Vector Mesons in Belle}

\author{A. Bo\.{z}ek \\
        for the Belle Collaboration}

\address{H. Niewodniczanski Institut of Nuclear Physics, Ul. Kawiory 26A, 
Krakow, PL 30-055, POLAND\\E-mail: bozek@belle2.ifj.edu.pl}


\twocolumn[\maketitle\abstracts{ 
Using the data sample of $10.5 \rm{fb}^{-1}$ collected by the Belle detector,
we searched for two-body charmless decays involving vector mesons.
The clear signal signature of  $B^{\pm} \to \phi K^{\pm}$ is seen and its branching fraction is measured to be $(1.39^{+.32}_{-.30} \pm 0.25) \times 10^{-5}$. 
The evidence for $ B^{\pm} \to \rho^{0} \pi^{\pm}$, $ B^{0} \to \rho^{\mp} \pi^{\pm}$ and  $B^{\pm} \to \phi K^{* \pm}$ is reported
and their branching fractions are determined. No significant signals are 
observed for $B^{0} \to \phi K^{0}_{S}$, $\to \phi K^{* 0}$, $\to \rho^{\mp} K^{\pm}$, $B^{\pm} \to \rho^{0} K^{\pm}$, 
$ \to \omega \pi^{\pm}$ and $ \to \omega K^{\pm}$
only the 90$\%$ C.L. upper limits are given.}]

\section{Introduction}

The study of charmless $B$ decays will play a
significant role in checking the KM matrix
description and in testing the validity
of the Standard Model description of CP violation.


Decays involving $b\!\! \to  ss\bar{s}$ transitions such as $B\!\!  \to   \phi K$ cannot occur 
via a tree process and
are expected to be dominated by the penguin diagram, thus their measurement
will give information on the strength of the penguin transition.
They are also sensitive to physics beyond the Standard Model,
since non-SM particles can contribute via additional loop diagrams.

The decay $B^0 \to \rho^{\pm} \pi^{\mp}$ is one of the preferred modes
for determining the angle $\phi_{2}$ of the unitarity triangle.
The decays  $ B^{\pm}\!\!  \to \rho^{0} \pi^{\pm}$ and $ B^{\pm}\!\!  \to \omega \pi^{\pm}$ are interesting as they
may exhibit large direct CP violation.
The  $ B^{\pm}\!\!  \to \rho^{0} K^{\pm}$ channel may yield information about the angle $\phi_{3}$.

The first searches for $B \to \phi K$~decays modes were performed by the CLEO
collaboration in data samples of gradually increasing integrated luminosities of
2.42, 3.11 and 
5.5 \rm{fb}$^{-1}$  but no evidence for these modes was found~\cite{cleo1999}.
The first evidence for the decay $B^{\pm}\!\!  \to \phi K^{\pm}$
in a data sample of $5.5 \times 10^{6} B\bar{B}$ pairs was reported by
the Belle collaboration at the ICHEP2000 Conference in Osaka\cite{conf07}.
The branching fraction was measured to be 
${\cal B}(B^{\pm} \!\! \to\!\!  \phi K^{\pm}) = ( 1.72^{+0.67}_{-0.54} \pm 0.18 ) 
\times 10^{-5}$. 
At the same conference, CLEO~\cite{cleo2000_conf} presented its first hints
of the observation of this decay mode, quoting a branching fraction of
$({0.64^{+0.25}_{-0.21}}^{+0.05}_{-0.20} ) \times 10^{-5}$.
Recently CLEO reported~\cite{clns0103} measurements based on the data sample of 
$9.7 \times 10^{6} B\bar{B}$ pairs:
${\cal B}(B^{\pm}\!\!  \to  \phi K^{\pm}) = ( 0.55^{+0.21}_{-0.18} 
\pm 0.06) \times 10^{-5}$ 
and 
${\cal B}(B^{0}\!\!  \to  \phi K^{* 0}) = ( {1.15^{+0.45}_{-0.37}}^{+0.18}_{-0.17} ) 
\times 10^{-5}$
and $90\%$ C.L. upper limits for the branching fractions of
$B^{* \pm}\!\!  \to \phi K^{* \pm}$ and $B^{0\!\! }\to \phi K^{0}$ decays of 
$2.25 \times 10^{-5}$ and $1.23 \times 10^{-5}$ respectively~\cite{clns0103}.

Charmless $B$ decays into modes with  $\rho (\pi/K)$ and $\omega (\pi/K)$
are hampered by large combinatorial backgrounds. These modes have only been seen 
recently, first by CLEO~\cite{cleo2000_rho} and then by BABAR~\cite{babar_rho}.
The current branching fractions measured by CLEO and BABAR are given  in Table~\ref{tab:broexp}.

\begin{table}
\begin{center}
\caption{The measured branching fractions by other experiments $[10^{-5}]$.}\label{tab:broexp}
\vspace{0.1cm}
\begin{tabular}{|c|c|c|} 
\hline 
\raisebox{0pt}[12pt][6pt]{Mode} & 
\raisebox{0pt}[12pt][6pt]{BABAR$(BF)$} & 
\raisebox{0pt}[12pt][6pt]{CLEO$(BF)$}  \\
\hline
\raisebox{0pt}[12pt][6pt]{$ B^{\pm}\!\! \to \rho^{0} \pi^{\pm}$} & 
\raisebox{0pt}[12pt][6pt]{$2.4 \pm 0.8 \pm 0.3$} & 
\raisebox{0pt}[12pt][6pt]{$1.04 \pm 0.33 \pm 0.21$} \\
\hline
\raisebox{0pt}[12pt][6pt]{$ B^{\pm}\!\! \to \rho^{0} K^{\pm}$} & 
\raisebox{0pt}[12pt][6pt]{$1.0 \pm 0.6 \pm 0.2$} & 
\raisebox{0pt}[12pt][6pt]{$0.84 \pm 0.38 \pm 0.18$} \\ 
\hline
\raisebox{0pt}[12pt][6pt]{$ B^{0}\!\! \to \rho^{\mp} \pi^{\pm}$} & 
\raisebox{0pt}[12pt][6pt]{$4.9 \pm 1.3^{+0.6}_{-0.5}$} & 
\raisebox{0pt}[12pt][6pt]{$2.76^{+0.84}_{-0.74} \pm 0.42$} \\
\hline
\raisebox{0pt}[12pt][6pt]{$ B^{0}\!\! \to \rho^{\mp} K^{\pm}$} & 
\raisebox{0pt}[12pt][6pt]{- } & 
\raisebox{0pt}[12pt][6pt]{$<3.3$ $(90\% C.L.)$ } \\ 
\hline
\raisebox{0pt}[12pt][6pt]{$ B^{\pm}\!\! \to \omega \pi^{\pm}$} & 
\raisebox{0pt}[12pt][6pt]{- } & 
\raisebox{0pt}[12pt][6pt]{$1.13 \pm 1.4$} \\
\hline
\raisebox{0pt}[12pt][6pt]{$ B^{\pm}\!\! \to \omega K^{\pm}$} & 
\raisebox{0pt}[12pt][6pt]{- } & 
\raisebox{0pt}[12pt][6pt]{$<0.79$ $(90\% C.L.)$ } \\ 
\hline
\raisebox{0pt}[12pt][6pt]{$ B^{\pm}\!\! \to \omega h^{\pm}$} & 
\raisebox{0pt}[12pt][6pt]{$0.89 \pm 0.54 \pm 0.22 $ } & 
\raisebox{0pt}[12pt][6pt]{- } \\
\hline
\end{tabular}
\end{center}
\end{table}
\vspace*{3pt}

\section{The data sample and the Belle detector}

The presented results are based on the data set collected by the Belle detector~\cite{belle_det} at KEKB~\cite{kekb}, the asymmetric B-factory at KEK.
It consists of $10.5 \rm{fb}^{-1}$ taken at the $\Upsilon(4S)$ resonance and $0.6 \rm{fb}^{-1}$ taken below 
the $B\bar{B}$ production threshold used for continuum studies.  
The Belle detector is a general purpose magnetic spectrometer equipped
with a 1.5~T superconducting solenoid magnet. 
Charged tracks are reconstructed in a 50 layer Central Drift Chamber (CDC) 
and in three concentric layers of double sided silicon strip detectors of the
Silicon Vertex Detector (SVD). The SVD allows precise reconstruction of 
secondary decay vertices.
The charged particle acceptance of the spectrometer covers the laboratory 
polar angles $ 17^{\circ} < \theta < 150^{\circ} $ which corresponds to~$\sim 90\%$
of the full CMS solid angle.
Photons and electrons are identified using the CsI(Tl) Electromagnetic Calorimeter (ECL) 
 located inside the magnet coil. 
Muons and $K^{0}_{L}$'s are detected using resistive plate
chambers embedded in the iron magnetic flux return (KLM).
Charged particles are identified using specific ionization losses
in the CDC and identification information from the Aerogel Cherenkov 
Counters (ACC) and Time of Flight Counters (TOF).
By these three 
methods, $K/\pi$ separation is achieved over the  momentum range from about
$0.2$ to $3.5 \,\rm GeV$.
The high momentum kaon identification efficiency is determined to be 
$\sim 80\%$ (at a purity of greater than $ 85\%$).

\section{Event selection}


In the search for the above channels we take full advantage of asymmetric 
$e^{+}e^{-}$ collisions resulting in boosted $B\bar{B}$ pairs, 
good vertexing and particle identification capabilities of 
the Belle detector.

The details of analysis are optimized for each channel separately.
For lack of space, only the main steps in the analysis can be
described.

We search for $\phi \to K^{+}K^{-}$ decays by selecting pairs of 
oppositely charged tracks consistent with  kaon hypothesis 
(which accepts kaons with $>90\%$ efficiency) 
and $| M_{KK} - M_{\phi}| <10 \rm{MeV}$. 
Events with a candidate $\phi$ are accepted if the
$\phi$ momentum in the CMS exceeds $2.0 \, \rm GeV$.

Candidate  $\rho^{0} \to \pi^{+} \pi^{-}$ decays are required to have
dipion invariant masses which satisfy 
 $| M_{\pi\pi} - M_{\rho^{0}}| <220 \rm{MeV}$, while the  $\rho^{\pm}$ candidates are required to satisfy
 $| M_{\pi^{\pm}\pi^{0}} - M_{\rho^{\pm}}| <200 \rm{MeV}$.

The $\omega$ is reconstructed in decay mode $\omega \to \pi^{+}\pi^{-}\pi^{0}$.
We require both $\pi^{0}$'s daughter photons 
carrying energy greater than $50 \rm{MeV}$. A variable, defined as 
the cross product of the $\pi^{+}\pi^{-}$ momentum vs $\pi^{0}$ in the $\omega$ rest frame (called  {\it $\omega$-amplitude} ) 
is used to suppress the combinatorial background.
The $\pi^{+}\pi^{-}\pi^{0}$ mass is required to be within a $\pm 30 \rm{MeV}$ window from the nominal $\omega$ mass. 

The charged tracks used in the $\omega/\rho$ reconstruction are required to be inconsistent with 
lepton or kaon. The required level of particle identification (PID)
for accompanying kaons $K^{\pm}$, $K^{* \pm}$, $K^{* 0}$,  
is optimized using the measured kaon efficiency and pion misidentification 
curves.
At the chosen cut value the kaon identification efficiency exceeded $80\%$.
For $K^{0}_{S} \to \pi^{+}\pi^{-}$ decays we use oppositely charged track pairs where the 
displacement of the $\pi^{+}\pi^{-}$ vertex from the interaction region in the
transverse $(r - \phi)$ plane is more than $1 \, \rm mm$. The $\phi$ 
coordinate of the $\pi^{+}\pi^{-}$ vertex point and the $\phi$ direction of 
the $\pi^{+}\pi^{-}$ vertex point agree within 0.2 radians and 
$|m_{\pi\pi}-m_{K^{0}}|< 15\, \rm \rm{MeV}$.
The $K^{*}$ candidates are reconstructed in four modes: $K^{* 0} \to K^{\pm} \pi^{\mp}$, 
$K^{* 0} \to K^{0}_{S} \pi^{0}$, $K^{* \pm} \to K^{\pm} \pi^{0}$ and $K^{* \pm} \to K^{0}_{S} \pi^{\pm}$.
The invariant mass is fitted with the beam spot constraint (described below) and the signal window
is defined within $\pm 50 \rm{MeV}$ of the nominal resonance mass.

\section{$B$ meson reconstruction}

The selected vector meson and the pseudoscalar (or vector) are combined to form the B-meson decay candidate.
The  vertex fit of the  $B$ candidate is  performed with a beam spot constraint 
($\sigma_{x} \approx 100 \,\mu {\rm m}$ , $\sigma_{y} \approx 5\, \mu {\rm m}$
and $\sigma_{z} \approx 3\, \rm mm$)
enlarged by a halo corresponding to the width expected for the $B$ meson 
lifetime ($\approx 20$ $\mu {\rm m}$ in the transverse plane).
A run dependent interaction point  position and beam spot 
size are used. 

\begin{figure}
\center
\psfig{figure=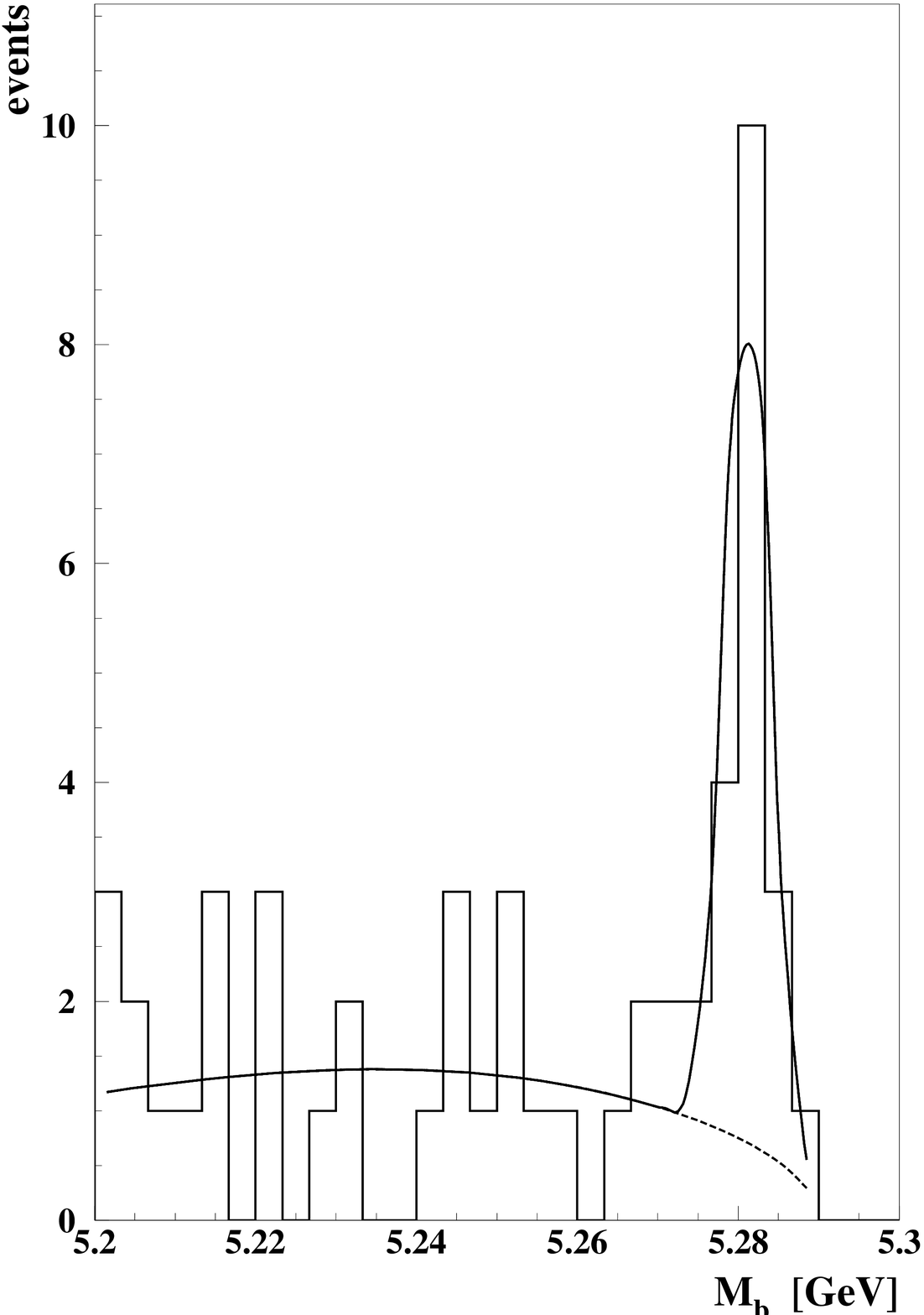,height=1.5in,width=3.6in}
\psfig{figure=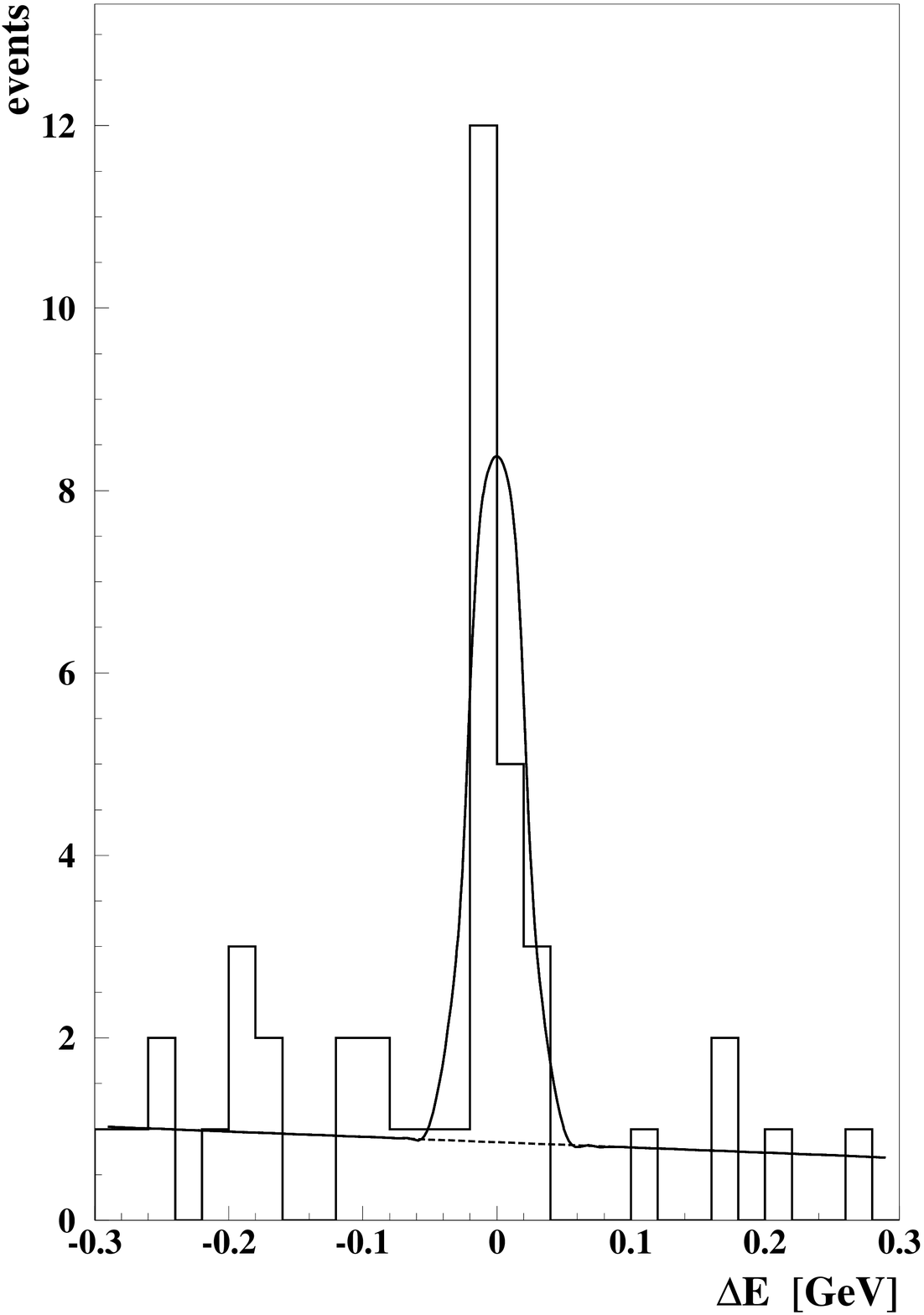,height=1.5in,width=3.6in}
\caption{The $M_{b}$ (top) and $\Delta E$ (bottom) distribution  for selected
$B^{\pm} \to \phi K^{\pm}$ candidates. The curves are projections of the unbinned ML fit.}
\label{fig:phikde}
\end{figure}

The candidate $B$ decay is identified by a beam constrained mass
$M_{b}=\sqrt{(\sqrt{s}/2)^2 - P^{* 2}_{B}}$ and the calculated energy difference $\Delta E = E^{*}_{B} -\sqrt{s}/2 $, where $E^{*}_{B}$ and $P^{* 2}_{B}$  are the energy and 3-momentum of the $B$ candidate in 
the $\Upsilon(4S)$ rest frame. The width of the $B$ signal region in $M_{b}$ and $\Delta E$ depends 
on the expected resolutions for each decay mode.

The continuum background has been suppressed using observables: 
the angle between the $B$ candidate momentum vector and 
the thrust vector of the remaining tracks of the event 
($|\cos (\theta_{thr-B})|$), 
and the $B$ candidate production angle in the CMS 
($|\cos (\theta^{*}_{B})|$).
For $B \to P V$ decay
channels an additional variable is used:
$|\cos ( \theta_{H} )|$, where the helicity angle 
$\theta_{H}$ is the angle between the direction of the $K/\pi$ and 
the momentum vector of $\phi/\rho/\omega$, in the vector meson rest frame
\footnote{In a Pseudoscalar($P$) $\to$ Pseudoscalar - Vector($V$) decay 
the vector meson is polarized, decaying to two pseudoscalars $ V \to PP$ according to a  $\cos^{2} ( \theta_{H} )$; the combinatorial background is flat in 
$\cos ( \theta_{H} )$.}.

\subsection{$B \to\phi K$}

\begin{figure}
\center
\psfig{figure=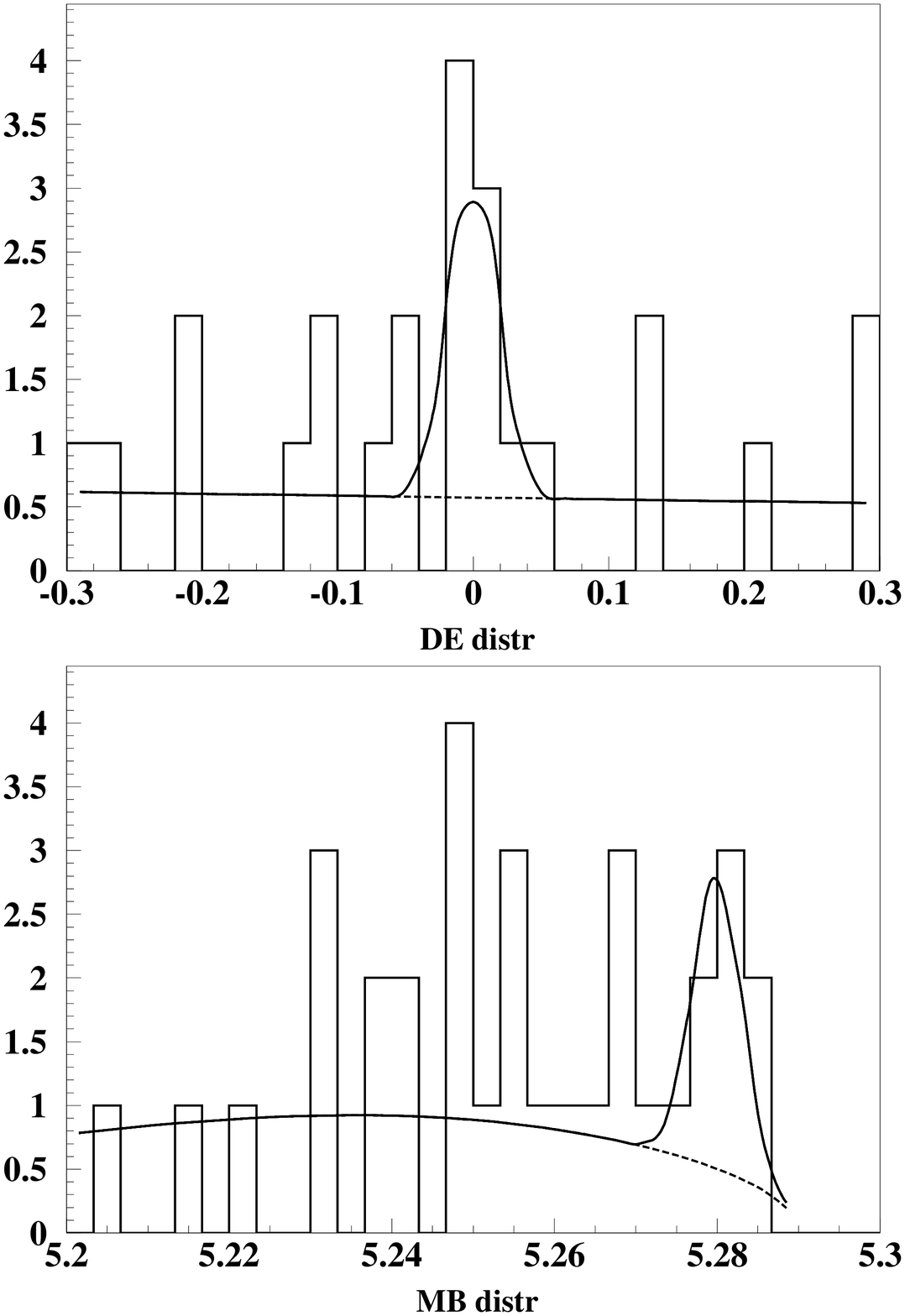,height=2.9in,width=3.6in}
\caption{The $M_{b}$ (top) and $\Delta E$ (bottom) distributions  for 
$B^{\pm} \to \phi K^{\pm *}$ candidates. The curves are projections of the unbinned ML fit.}
\label{fig:phikstde}
\end{figure}

\begin{table}
\begin{center}
\caption{The fit results  for $B \to \phi K$ modes.}\label{tab:fits}
\vspace{0.1cm}
\begin{tabular}{|c|c|c|} 
\hline 
\raisebox{0pt}[12pt][6pt]{Mode} & 
\raisebox{0pt}[12pt][6pt]{fitted signal yield} & 
\raisebox{0pt}[12pt][6pt]{significance}  \\
\hline
\raisebox{0pt}[12pt][6pt]{$B^{\pm} \to \phi K^{\pm}$} & 
\raisebox{0pt}[12pt][6pt]{$17.8^{+4.8}_{-4.2}$} & 
\raisebox{0pt}[12pt][6pt]{$7.5 \sigma$} \\
\hline
\raisebox{0pt}[12pt][6pt]{$B^{0} \to \phi K^{* 0}$}& 
\raisebox{0pt}[12pt][6pt]{$6.5^{+3.5}_{-2.7}$} &  
\raisebox{0pt}[12pt][6pt]{$3.6 \sigma$} \\
\hline
\raisebox{0pt}[12pt][6pt]{$B^{\pm} \to \phi K^{* \pm}$} & 
\raisebox{0pt}[12pt][6pt]{$1.3^{+1.8}_{-1.1}$} & 
\raisebox{0pt}[12pt][6pt]{$1.4 \sigma$} \\
\hline
\raisebox{0pt}[12pt][6pt]{$B^{0} \to \phi K^{0}_{S}$}& 
\raisebox{0pt}[12pt][6pt]{$0.9^{+1.7}_{-0.9}$} &  
\raisebox{0pt}[12pt][6pt]{$1.2 \sigma$} \\
\hline
\end{tabular}
\end{center}
\end{table}
\vspace*{3pt}

The continuum background suppression is performed by sequential cuts;
$|\cos (\theta_{thr-B})|<0.8$ and $|\cos (\theta^{*}_{B})|<0.8$.
For $B^{\pm} \to \phi K^{\pm}$ and $B^{0} \to \phi K^{0}_{S}$
channels an additional cut is used:
$|\cos ( \theta_{H} )|>0.5$.
The signal yield in each studied decay channel 
is  extracted from the unbinned maximum-likelihood fit.
The fits are made simultaneously in $M_{b}$ - $\Delta E$ plane.
The signal in $M_{b}$ and $ \Delta E$ is adequately represented by a single Gaussian.
The background distribution for $M_{b}$ is represented by the ARGUS 
function\footnote{
$B(x)= N \sqrt{1-x^{2}}\exp(P(1-x^{2}))$; where 
$x = \frac{M_{b}}{E^{*}_{beam}}$, $P$ is the shape variable fitted 
to background sample and N is the overall normalization factor fitted 
to the distribution in the signal region.}.
The background in $\Delta E$ is  parametrized by a linear function.

The accepted candidates for $B^{\pm} \to \phi K^{\pm}$ are shown in
Fig.~\ref{fig:phikde}. The fit gives 
 $17.8^{+4.8}_{-4.2}$ signal events. The statistical significance of the signal yield 
is $7.5 \sigma $\footnote{The statistical significance of the signal yield is 
defined  as 
$\sqrt{-2\log({L_{0}/L_{max}})}$, where
$L_{0}$  corresponds to the value at zero signal yield hypothesis.}.
The efficiency for this decay mode is $11.8 \%$, which includes 
detector acceptance and the intermediate BF.

The results of the fit for $B^{\pm} \to \phi K^{* \pm}$ decays candidates is shown in 
Fig.~\ref{fig:phikstde} . The fitted number of signal events is $6.5^{+3.5}_{-2.7}$ with significance of $3.6 \sigma$ .
The efficiency for this decay mode is $4.0 \%$.

The results of the fits for other $B \to \phi K$ modes are summarized in Table~\ref{tab:fits}.

\subsection{$ B \to \rho \pi^{\pm}$}

In these channels another discrimination observables are used: an angle between  the $\rho$ and the beam axis in the $B$ rest frame.
In addition the extended Fox-Wolfram\cite{sfw} moments are used (Super Fox-Wolfram moments ).
The moments are combined to maximize the separation
between signal and background distributions.

The variables are combined into  
Likelihood ratio ${\cal L}$ for better background rejection.
The  probabilities of being the signal ($p^{sig}$) 
and background ($p^{bck}$) are calculated for each of the variables and the likelihood ratio is formed 
${\cal L}= \frac{p^{sig}}{p^{sig}+p^{bck}}$. A cut is made at ${\cal L} >0.9$.

The possible cross talk from $B^{\pm} \to D^{0}\pi^{\pm}; D^{0} \to K^{\pm} \pi^{\mp}$
which can imitate $B^{\pm} \to \rho^{0}\pi^{\pm}; \rho \to \pi^{+} \pi^{-}$ 
is removed by checking the $D^{0}$ decay hypothesis.

The fit for $B^{\pm} \to \rho^{0}\pi^{\pm}$ in the $\Delta E$  projection only (with $M_{b}$ cut) gives the 
signal yield of $13.73^{+6.53}_{-5.85}$ at significance of $2.9 \sigma$
 and efficiency of $13.4 \%$.
The result is confirmed in the fit to the $M_{b}$  distribution, which gives a significance of $3.4 \sigma$.


 \begin{table}
\begin{center}
\caption{The fit results  for $B^{0} \to \rho^{\pm} \pi^{\mp}$ modes.}\label{tab:fitsrho}
\vspace{0.1cm}
\begin{tabular}{|c|c|c|} 
\hline 
\raisebox{0pt}[12pt][6pt]{data sample} & 
\raisebox{0pt}[12pt][6pt]{fitted signal yield} & 
\raisebox{0pt}[12pt][6pt]{significance}  \\
\hline
\raisebox{0pt}[12pt][6pt]{$\cos ( \theta_{H} )>0$} & 
\raisebox{0pt}[12pt][6pt]{$10.9^{+5.5}_{-4.8}$} & 
\raisebox{0pt}[12pt][6pt]{$2.6 \sigma$} \\
\hline
\raisebox{0pt}[12pt][6pt]{$\cos ( \theta_{H} )<0$}& 
\raisebox{0pt}[12pt][6pt]{$10.1^{+5.5}_{-4.7}$} &  
\raisebox{0pt}[12pt][6pt]{$2.4 \sigma$} \\
\hline
\end{tabular}
\end{center}
\end{table}
\vspace*{3pt}

For the  $B^{0} \to \rho^{\pm}\pi^{\mp}$ mode,
the $\cos ( \theta_{H} ) >0$ and 
$\cos ( \theta_{H} ) < 0$ regions are fitted separately.
The yield is determined from a simultaneous fit to the 
$M_{b}$ and $\Delta E$ distributions

The fit results are shown in the Table~\ref{tab:fitsrho}.
The significance of the signal yield predicted by the fit to both subsamples combined is  $3.5 \sigma$. 

\subsection{$ B \to \rho K^{\pm}$}

For the  $B^{\pm} \to \rho^{0} K^{\pm}$ mode the fit
is made simultaneously in the $M_{b}$ and $\Delta E$ planes.
The fit results are given in the Table~\ref{tab:fitsrhok}.
The significance of the signal is only $2.2 \sigma$.

 \begin{table}
\begin{center}
\caption{The fit results for $B \to \rho^{\pm} K^{\mp}$ modes.}\label{tab:fitsrhok}
\vspace{0.1cm}
\begin{tabular}{|c|c|c|} 
\hline 
\raisebox{0pt}[12pt][6pt]{data sample} & 
\raisebox{0pt}[12pt][6pt]{fitted signal yield} & 
\raisebox{0pt}[12pt][6pt]{significance}  \\
\hline
\raisebox{0pt}[12pt][6pt]{$\cos ( \theta_{H} )>0$} & 
\raisebox{0pt}[12pt][6pt]{$5.0^{+3.8}_{-3.3}$} & 
\raisebox{0pt}[12pt][6pt]{$1.9 \sigma$} \\
\hline
\raisebox{0pt}[12pt][6pt]{$\cos ( \theta_{H} )<0$}& 
\raisebox{0pt}[12pt][6pt]{$4.6^{+4.1}_{-3.3}$} &  
\raisebox{0pt}[12pt][6pt]{$1.5 \sigma$} \\
\hline
\end{tabular}
\end{center}
\end{table}
\vspace*{3pt}

For the $B^{0} \to \rho^{\pm} K^{\mp}$ mode  the fit in 
the $\Delta E$  projection is performed  (with $M_{b}$ cut). The fit gives
signal yield of $8.29^{+5.07}_{-4.43}$ with significance of $2.0 \sigma$
(in the case of the fit to $M_{b}$ only the significance is $1.8 \sigma$) and efficiency of
$12.8 \%$. 

\subsection{$ B^{\pm} \to \omega h^{\pm}$}

As in the $ B \to \rho h$ search, a cut on a likelihood ratio which includes the
{\it $\omega$-amplitude}is used
to suppress the continuum background. 
The cut is set at ${\cal L} >0.95$. 
The unbinned likelihood fit to $M_{b}$ and $\Delta E$ is used to extract the signal yields. 
No signal is observed.

\begin{table}
\begin{center}
\caption{The fit results for $B \to \omega h$ modes.}\label{tab:fitsom}
\vspace{0.1cm}
\begin{tabular}{|c|c|c|} 
\hline 
\raisebox{0pt}[12pt][6pt]{Mode} & 
\raisebox{0pt}[12pt][6pt]{fitted signal yield} & 
\raisebox{0pt}[12pt][6pt]{significance}  \\
\hline
\raisebox{0pt}[12pt][6pt]{$B^{\pm} \to \omega \pi^{\pm}$} & 
\raisebox{0pt}[12pt][6pt]{$2.8^{+3.6}_{-2.7}$} & 
\raisebox{0pt}[12pt][6pt]{$1.0 \sigma$} \\
\hline
\raisebox{0pt}[12pt][6pt]{$B^{\pm} \to \omega K^{\pm}$}& 
\raisebox{0pt}[12pt][6pt]{$3.4^{+3.0}_{-2.2}$} &  
\raisebox{0pt}[12pt][6pt]{$1.8 \sigma$} \\
\hline
\raisebox{0pt}[12pt][6pt]{$B^{\pm} \to \omega h^{\pm}$}& 
\raisebox{0pt}[12pt][6pt]{$6.4 \pm8.9$} &  
\raisebox{0pt}[12pt][6pt]{$1.8 \sigma$} \\
\hline
\end{tabular}
\end{center}
\end{table}
\vspace*{3pt}

The $\omega \pi$ and $\omega K$ yields are obtained from the simultaneous fit. For the $\omega h$ mode, the particle identification information is not applied in the fast charged tracks selection.  The results are summarized in Table~\ref{tab:fitsom}.

\section{Conclusions}

Using the data sample of $10.5 \rm{fb}^{-1}$ collected by the Belle detector,
a search was performed for two-body charmless decays including vector meson.
A clear signal signature of  $B^{\pm} \to \phi K^{\pm}$ is seen with 
$7.5 \sigma$  significance. The branching fraction for  this mode is measured to be $(1.39^{+.32}_{-.30} \pm 0.25) \times 10^{-5}$. 
Evidence  $B^{\pm} \to \rho^{0} \pi^{\pm}$, $ B^{0} \to \rho^{\mp} \pi^{\pm}$ and  $B^{\pm} \to \phi K^{* \pm}$ is reported.
For $B^{0} \to \phi K^{0}_{S}$, $B^{0} \to \phi K^{* 0}$,
$ B^{\pm} \to \rho^{0} K^{\pm}$, $ B^{0} \to \rho^{\mp} K^{\pm}$, 
$ B^{\pm} \to \omega \pi^{\pm}$ and $ B^{\pm} \to \omega K^{\pm}$
 no significant signals are observed and the 90$\%$ C.L. limits are
given.

\begin{table}
\begin{center}
\caption{The summary of the obtained branching fractions and upper limits 
[$\times 10^{-5}$]. }\label{tab:BF}
\vspace{0.1cm}
\begin{tabular}{|c|c|c|} 
\hline 
\raisebox{0pt}[12pt][6pt]{Mode} & 
\raisebox{0pt}[12pt][6pt]{BF} & 
\raisebox{0pt}[12pt][6pt]{UL $90 \%$ C.L.}  \\
\hline
\raisebox{0pt}[12pt][6pt]{$B^{+} \to \phi K^{+}$} & 
\raisebox{0pt}[12pt][6pt]{${1.39^{+0.37}_{-0.33}}^{+0.14}_{-0.24}$} & 
\raisebox{0pt}[12pt][6pt]{-} \\
\hline
\raisebox{0pt}[12pt][6pt]{$B^{0} \to \phi K^{0}$}& 
\raisebox{0pt}[12pt][6pt]{-} &  
\raisebox{0pt}[12pt][6pt]{$< 1.6$} \\
\hline
\raisebox{0pt}[12pt][6pt]{$B^{0} \to \phi K^{*0}$}& 
\raisebox{0pt}[12pt][6pt]{$1.5^{+0.8}_{-0.6} \pm 0.3$} &  
\raisebox{0pt}[12pt][6pt]{$<3.0$ } \\
\hline
\raisebox{0pt}[12pt][6pt]{$B^{+} \to \phi K^{*+}$}& 
\raisebox{0pt}[12pt][6pt]{$-$} &  
\raisebox{0pt}[12pt][6pt]{$<  3.6$} \\
\hline
\raisebox{0pt}[12pt][6pt]{$B^{+} \to \rho^0\pi^+$}& 
\raisebox{0pt}[12pt][6pt]{$1.12^{+0.53}_{-0.48} \pm 0.19$} &  
\raisebox{0pt}[12pt][6pt]{$< 2.88$} \\
\hline
\raisebox{0pt}[12pt][6pt]{$B^{+} \to \rho^0 K^+ $ }& 
\raisebox{0pt}[12pt][6pt]{$-$} &  
\raisebox{0pt}[12pt][6pt]{$< 1.35$} \\
\hline
\raisebox{0pt}[12pt][6pt]{$B^{0} \to \rho^+\pi^-$}& 
\raisebox{0pt}[12pt][6pt]{$2.02^{+0.83}_{-0.66} \pm 0.33 $} &  
\raisebox{0pt}[12pt][6pt]{$< 3.57$} \\
\hline
\raisebox{0pt}[12pt][6pt]{$B^{0} \to \rho^+ K^- $}& 
\raisebox{0pt}[12pt][6pt]{$-$} &  
\raisebox{0pt}[12pt][6pt]{$< 2.36$} \\
\hline
\raisebox{0pt}[12pt][6pt]{$B^{+} \to \omega h^+$}& 
\raisebox{0pt}[12pt][6pt]{$-$} &  
\raisebox{0pt}[12pt][6pt]{$< 1.41$} \\
\hline
\raisebox{0pt}[12pt][6pt]{$B^{+} \to \omega\pi^+$}& 
\raisebox{0pt}[12pt][6pt]{$-$} &  
\raisebox{0pt}[12pt][6pt]{$<  0.94$} \\
\hline
\raisebox{0pt}[12pt][6pt]{$B^{+} \to \omega K^+$}& 
\raisebox{0pt}[12pt][6pt]{$-$} &  
\raisebox{0pt}[12pt][6pt]{$< 1.05$} \\
\hline
\end{tabular}
\end{center}
\end{table}

\section*{Acknowledgments}
We gratefully 
acknowledge support from 
the Polish State Committee for Scientific Research 
under contract No.2P03B 17017.

\end{document}